\documentclass[11pt]{article}
\usepackage{a4wide}

\usepackage{amsmath}
\usepackage{bm}
\usepackage{graphicx}
\usepackage{subfig}
\usepackage[square, comma, sort&compress]{natbib}
\newcommand{\SU}[1]{\mathrm{SU}(#1)}

\newcommand{\Tr}[2]{\mathop{\mathrm{Tr}}_{#1}\left[#2\right]}
\newcommand{\pdiff}[2]{\frac{\partial#1}{\partial#2}}

\newcommand{\nbrack}[1]{\left(#1\right)}
\newcommand{\sbrack}[1]{\left[#1\right]}

\newcommand{\be}{\begin{equation}}
\newcommand{\ee}{\end{equation}}
\newcommand{\ba}{\begin{eqnarray}}
\newcommand{\ea}{\end{eqnarray}}

\begin{document}

\date{\small{\em Presentation at the XLIVth Rencontres de Moriond on Electroweak Interactions}\\
\small{\em and Unified Theories, La Thuile, Italy, 7-14 Mar 2009}}
\title{\bf{SUSY Gauge Singlets and Dualities}}
\author{James Barnard\\[3ex]
\small{\em Department of Mathematics, Durham University, Durham DH1 3LE, UK}\\[1.5ex] 
\small{james.barnard@durham.ac.uk}}
\maketitle
\abstract{\noindent We discuss how, by including gauge singlets in supersymmetric gauge theories, one can construct and test new types of duality.  This may help in finding dual theories of supersymmetric GUTs.  This talk is based on the recent article Ref.\cite{Abel:2009ty}.}

\section{Introduction and motivation}

\noindent Seiberg duality in $\mathcal{N}=1$ supersymmetry (see Ref.\cite{Intriligator:1995au} for a review) gives us a different perspective on supersymmetric (SUSY) gauge theories.  It is a pure field theory duality between two SUSY gauge theories.  The original incarnation was supersymmetric QCD (SQCD) with $F_Q$ flavours of quark/antiquark and $N$ colours.  The would-be dual theory is also SQCD, with $F_Q$ flavours of quark/antiquark, but with $n=F_Q-N$ colours and an extra, elementary meson field $M$. Various powerful tests such as `t Hooft anomaly matching established that the two theories are really different descriptions of the same infra-red physics.  This kind of duality has great potential to shed light on many aspects of BSM physics such as gauge unification, proton decay and dynamical SUSY breaking.  Unfortunately Seiberg dualities only currently exist for theories with highly constrained matter content and unrealistic superpotentials (see Refs.\cite{Kutasov:1995ve,Kutasov:1995np,Kutasov:1995ss,Brodie:1996vx,Brodie:1996xm,Intriligator:1995ax} for some examples).  If we ever want to access these phenomenological applications we will therefore need to extend the idea of Seiberg duality to a more realistic model.  Ideally we would like to be able to find a duality involving a SUSY grand unified theory (GUT), like one of the $\SU{5}$ models.

\subsection{An example: ``dualification''}

\noindent To illustrate what Seiberg duality might do for us, consider a recent example from Ref.\cite{Abel:2008tx}; ``dualification''.  Suppose we have some SUSY breaking GUT where the SUSY breaking is mediated directly.  The set-up for direct mediation is as follows.  The theory contains two sectors; the visible sector and the hidden sector.  SUSY is broken somehow in the hidden sector then communicated to the visible sector by messenger particles.  For direct mediation, these messenger particles are charged under the visible sector gauge group.

\begin{figure}[!h]
\begin{center}
\includegraphics[width=10cm]{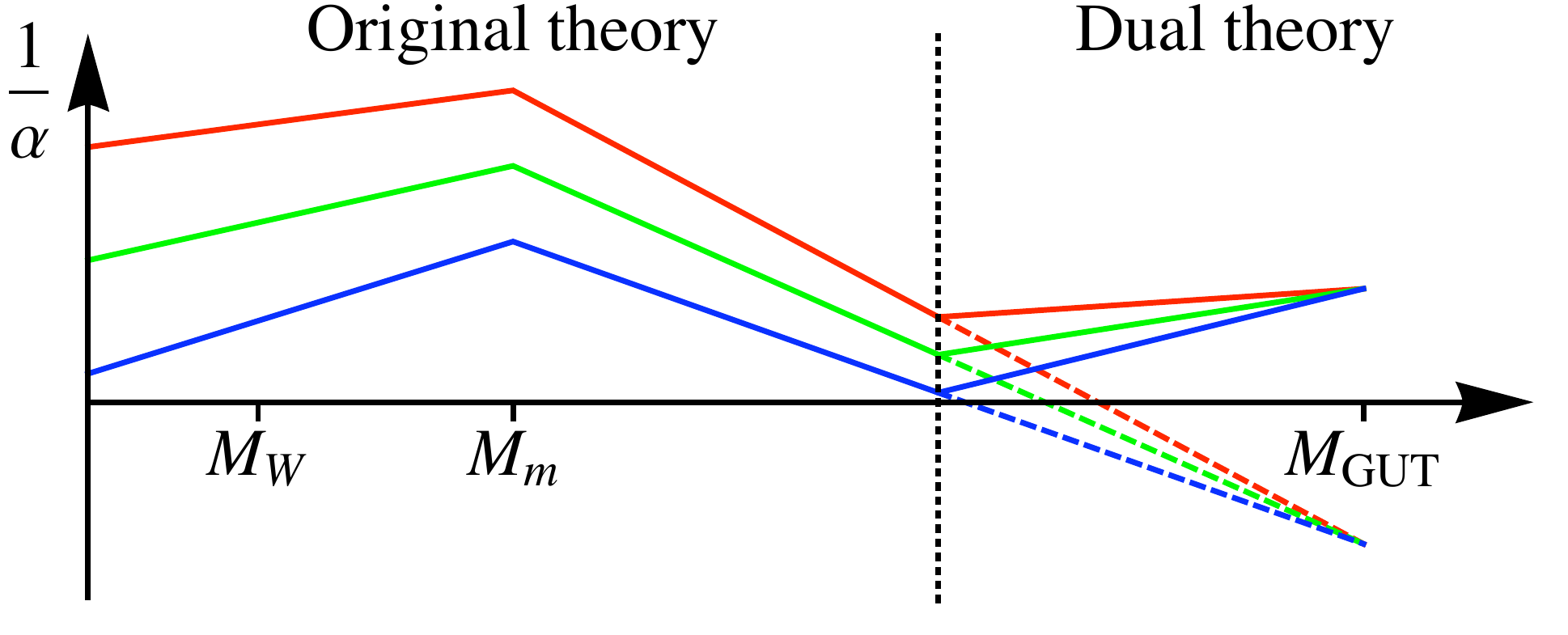}
\caption{\em ``Dualification''.  Deflection of the RG flows by messenger particles at the scale $M_m$ appears to cause unphysical unification at $1/\alpha<0$ in the original theory.  In the dual theory, unification occurs much more naturally.\label{fig:dualification}}
\end{center}
\end{figure}

Now consider the renormalisation group (RG) flow of the visible sector gauge couplings in this theory.  They are sketched in Figure \ref{fig:dualification} (where we show the inverse coupling $1/\alpha$).  Starting at the weak scale $M_W$ it looks initially as though the couplings in the theory are going to unify at a physical value of $1/\alpha$ at the scale $M_{\mathrm{GUT}}$.  However, once we reach the messenger scale $M_m$ we have to include the effects of the messenger particles, and thus the RG flows of the couplings are deflected.  This only occurs because the messenger particles are charged under the visible sector gauge group.  If the messenger sector comes in complete $SU(5)$ multiplets, all the beta functions are deflected by the same amount so the theory still unifies at the scale $M_{\mathrm{GUT}}$.  However, if the RG flows are deflected too much the theory encounters a Landau pole; the couplings become arbitrarily strong and hit $1/\alpha=0$ below the unification scale, where our description of the physics breaks down.  An interesting point to note is that the theory continues to look as though it unifies at the GUT scale, but at a negative (unphysical) value of $1/\alpha$.

Instead, we consider the original theory as one half of a duality.  When we reach the strong coupling regime $1/\alpha\approx0$ we can no longer describe the physics with the original theory but it may be that a dual description exists.  In the dual theory, the RG flow proceeds in the opposite direction so we move away from the Landau pole.  Remarkably, the unification is not affected by the duality.  All gauge couplings still flow to a single value and this occurs at the same scale as the original theory, $M_{\mathrm{GUT}}$.  It was therefore proposed in Ref.\cite{Abel:2008tx} that unphysical, negative gauge coupling unification in a SUSY gauge theory is actually a remnant from physical unification in a dual theory.

\section{Extending the duality}

\noindent Ref.\cite{Abel:2008tx} showed that this happens in simple vector-like GUT theories based on so-called Kutasov dual theories (of which more in a moment): alas no similar dual theories are known for the Georgi-Glashow model or Flipped $\SU{5}$.  If one could be found however, the same arguments should apply.  In order to do this various extensions of Seiberg duality are required, including chirality, more generations and so on.  The present 
work constitutes a step in this direction.

The difficulty in extending Seiberg duality stems from the meson sector.  One of the standard tests for dualities is that the classical moduli spaces of the two theories match up.  This means that all gauge invariant degrees of freedom in the original theory must have a counterpart in the dual theory, with exactly the same properties under all global symmetries.  In particular, there must be a one to one correspondence between mesons in the two theories.  As an example, consider regular Seiberg duality \cite{Intriligator:1995au}.

The mesons in the original theory are the usual SQCD composite mesons $\tilde{Q}Q$, where $Q$ and $\tilde{Q}$ are the quarks/antiquarks respectively.  The flavour indices have been suppressed and the colour indices summed over.  In the dual theory there are the composite mesons $\tilde{q}q$, where $q$ and $\tilde{q}$ are the dual quarks/antiquarks, but also the elementary mesons $M$.  We can see that there are twice as many mesons in the dual theory than in the original theory.  To remedy this we add a superpotential $W_{\mathrm{dual}}=M\tilde{q}q$ to the dual theory which projects out the composite mesons via the $F$-term equation
\be
\pdiff{W_{\mathrm{dual}}}{M}=\tilde{q}q=0.
\ee
We are now able to construct the map $\tilde{Q}Q\leftrightarrow M$ between mesons in the original theory and the dual theory.  This map respects all global symmetries.

For phenomenological purposes we often need to include more complicated matter than quarks and antiquarks (which live in the fundamental and antifundamental representations of the gauge group respectively).  For example, suppose we want to include an adjoint representation of the gauge group $X$.  The most general meson we can construct is now $\tilde{Q}X^jQ$ for \emph{any} positive integer $j$.  There are thus an arbitrary number of mesons in the original theory so the dual theory is not well defined (in particular, we would require infinitely many elementary degrees of freedom and an infinite number of colours in the dual theory to match the two theories).

The solution to this problem was discovered in Refs.\cite{Kutasov:1995ve,Kutasov:1995np,Kutasov:1995ss}.  The idea is to add a superpotential to the original theory which \emph{truncates the chiral ring}, in effect limiting the number of mesons in the original theory.  This can be accomplished with the superpotential $W_{\mathrm{orig}}=\Tr{}{X^{k+1}}$ for some integer $k$, where the trace is taken over colour indices.  The $F$-term equations in the original theory set
\be
\pdiff{W_{\mathrm{orig}}}{X}\sim X^k=0
\ee
so we can only build $k$ mesons: $\tilde{Q}X^jQ$ for $j=0,\ldots,k-1$.  Having done this, a duality can be found in a similar way to the original Seiberg duality, but with $n=kF_Q-N$ colours and a more complicated dual superpotential.

Unfortunately this technique comes with its own problems.  By adding a superpotential to the original theory we reduce the number of global symmetries.  Non-trivial global symmetries are crucial for testing the duality at a quantum mechanical level via highly non-trivial 't Hooft anomaly matching conditions.  If we add too many terms to the original theory's superpotential we will therefore be unable to rigourously test the duality.  Priority in this regard is given to $R$-symmetries; symmetries which do not commute with SUSY transformations (i.e.\ fermions have different charges to their scalar superpartners).  It turns out that $R$-symmetries generally give the most stringent test of the duality, so we must aim to choose a superpotential which allows one.

To address this problem we proposed the following in Ref.\cite{Abel:2009ty}.  If we add gauge singlets to the theory we can use them to restore the global symmetries broken by the original theory's superpotential.  Alternatively this process can be viewed as elevating the coupling constants in the superpotential to fields.  In particular, gauge singlets allow the preservation of an $R$-symmetry.  This allows us much more freedom in choosing the superpotential while preserving access to all of the standard tests of duality.  Now we have more freedom in choosing the superpotential we are able to add more matter to the theory and still have well behaved mesons.  We used this technique to extend Seiberg duality to theories with matter content closer to that of a SUSY GUT; multiple generations of adjoint or antisymmetric representations of the gauge group.

\subsection{Example: SQCD with three generations of antisymmetric tensor}

\noindent Suppose we want to extend SQCD (with $F_Q$ flavours of quark/antiquark and $N$ colours) to include three generations of antisymmetric tensor $A$, $B$ and $C$ (plus their conjugates).  Models with a single generation of antisymmetric were first discussed in Ref.\cite{Intriligator:1995ax} but for a realistic $\SU{5}$ GUT we need at least three.  To do this we include a singlet $\phi$ and give the original theory the superpotential
\be
W_{\mathrm{orig}}=\phi^{\rho_A}(A\tilde{A})^{k_A+1}+\phi^{\rho_B}(B\tilde{B})^{k_B+1}+\phi^{\rho_C}(C\tilde{C})^{k_C+1}+\phi^{\sigma}\nbrack{A\tilde{B}+\tilde{A}B+B\tilde{C}+\tilde{B}C}.
\ee
The $k$'s are positive integers and the $\rho$'s are to be determined.  The addition of a singlet is necessary for the theory to possess an $R$-symmetry.  For non-zero $\phi$, the $F$-terms of this superpotential result in truncation equations
\ba
\tilde{A}(A\tilde{A})^{k^*}&\sim&O\nbrack{\tilde{A}(A\tilde{A})^{k^*-2k_Ak_B-k_A-k_B}}\nonumber\\
(A\tilde{A})^{k^*}A&\sim&O\nbrack{(A\tilde{A})^{k^*-2k_Ak_B-k_A-k_B}A}.
\ea
for the chiral ring, where
\be
k^*=\frac{1}{2}\sbrack{\nbrack{2k_A+1}\nbrack{2k_B+1}\nbrack{2k_C+1}-1}.
\ee
The other antisymmetrics can be expressed entirely in terms of $A$ and $\tilde{A}$ so the truncation is complete.  The mesons are then schematically
\be
\begin{array}{lclcl}
M_j&\sim&\tilde{Q}(\tilde{A}A)^jQ,\hspace{5mm}&&j=0,\ldots,k^*\\
P_j&\sim&Q(\tilde{A}A)^j\tilde{A}Q,\hspace{5mm}&&j=0,\ldots,k^*-1\\
\tilde{P}_j&\sim&\tilde{Q}A(\tilde{A}A)^j\tilde{Q},\hspace{5mm}&&j=0,\ldots,k^*-1.
\end{array}
\ee

In Ref.\cite{Abel:2009ty} we show that a dual theory with
\be
n=\nbrack{2k^*+1}F_Q-4k^*-N
\ee
colours can then be found.  A key feature of our technique is that the gauge singlet appears in the definition of the mesons in the original theory.  We determine unambiguously how this should be done and find that our approach is totally consistent with all global symmetries as well as 't Hooft anomaly matching.

\section*{Acknowledgements}

\noindent I would like to thank Steven Abel for collaboration on this work and am grateful to the STFC and the meeting organizers for financial support.

\end{document}